\documentclass[conference]{IEEEtran}
\IEEEoverridecommandlockouts
\usepackage{cite}
\usepackage{amsmath,amssymb,amsfonts}
\usepackage{algorithmic}
\usepackage{graphicx}
\usepackage{textcomp}
\usepackage{xcolor}
\def\BibTeX{{\rm B\kern-.05em{\sc i\kern-.025em b}\kern-.08em
    T\kern-.1667em\lower.7ex\hbox{E}\kern-.125emX}}

\usepackage{url}
\usepackage{siunitx}
\usepackage{subcaption}
\usepackage[acronym]{glossaries}
\makeglossaries
\newacronym{ofdm}{OFDM}{Orthogonal Frequency Division Multiplexing}
\newacronym{mimo}{MIMO}{Multiple-Input Multiple-Output}
\newacronym{crlb}{CRLB}{Cramér-Rao Lower Bound}
\newacronym[plural=SAFs]{saf}{SAF}{Spatial Ambiguity Function}
\newacronym{csi}{CSI}{Channel State Information}
\newacronym{los}{LoS}{Line-of-Sight}
\newacronym{ue}{UE}{User Equipment}
\newacronym{cfo}{CFO}{Carrier Frequency Offset}
\newacronym{ula}{ULA}{Uniform Linear Array}
\newacronym{pmsr}{PMSR}{Peak-to-Median Sidelobe Ratio}
\newacronym{fim}{FIM}{Fisher Information Matrix}
\newacronym{efim}{EFIM}{Effective FIM}
\newacronym{rmse}{RMSE}{Root Mean Squared Error}
\newacronym{lse}{LSE}{Least Squares Estimator}
\newacronym{isac}{ISAC}{Integrated Sensing and Communication}
\newacronym{mocap}{MoCap}{Motion Capture}

\usepackage[font=footnotesize,labelfont=bf]{caption}
\begin{document}

\title{
Robust Localization in OFDM-Based Massive MIMO through Phase Offset Calibration
\thanks{This work is supported by the EU-HORIZON-MSCA-2022-DN No. 101119652 (6\textsuperscript{th}Sense), Horizon Europe Research and Innovation programme No. 101192521 (MultiX), and No. 101139257 (SUNRISE-6G).
}
}

\author{
  Qing Zhang\textsuperscript{*}, 
  Adham Sakhnini\textsuperscript{*,\dag},
  Robbert Beerten\textsuperscript{*},
  Haoqiu Xiong\textsuperscript{*}, \\
  Zhuangzhuang Cui\textsuperscript{*},
  Yang Miao\textsuperscript{\#}, 
  and Sofie Pollin\textsuperscript{*,\dag} \\
  \textsuperscript{*}Department of Electrical Engineering (ESAT), KU Leuven, Belgium \\
  \textsuperscript{\#}Department of Electrical Engineering (EEMCS-EE), University of Twente, The Netherlands \\
  \textsuperscript{\dag}Interuniversity Microelectronics Centre (IMEC), Leuven, Belgium
  \vspace{-0.2cm}

}

\maketitle

\begin{abstract}
Accurate localization in \gls{ofdm}-based massive \gls{mimo} systems depends critically on phase coherence across subcarriers and antennas. However, practical systems suffer from frequency-dependent and (spatial) antenna-dependent phase offsets, degrading localization accuracy. This paper analytically studies the impact of phase incoherence on localization performance under a static \gls{ue} and \gls{los} scenario. We use two complementary tools. First, we derive the \gls{crlb} to quantify the theoretical limits under phase offsets. Then, we develop a \gls{saf}-based model to characterize ambiguity patterns. Simulation results reveal that spatial phase offsets severely degrade localization performance, while frequency phase offsets have a minor effect in the considered system configuration. To address this, we propose a robust \gls{csi} calibration framework and validate it using real-world measurements from a practical massive \gls{mimo} testbed. The experimental results confirm that the proposed calibration framework significantly improves the localization \gls{rmse} from \SI{5}{m} to \SI{1.2}{cm}, aligning well with the theoretical predictions. 
\end{abstract}


\section{Introduction}

The rapid evolution of wireless communication systems has led to the widespread adoption of \acrfull{ofdm} in conjunction with massive \acrfull{mimo} architectures. These systems promise high spatial resolution and corresponding robust localization, which empowers \gls{isac} for next-generation wireless networks \cite{6g_white_paper, radio_localization, position_orientation}. However, a critical issue has arisen from imperfect phase coherence across subcarriers and antenna elements due to hardware impairments. 

In practice, \gls{ofdm} systems inherently suffer from frequency-dependent phase distortions arising from hardware impairments such as oscillator drift, \gls{cfo}, and IQ imbalance. Throughout the spatial dimension, antenna-dependent phase offsets due to hardware heterogeneity also lead to a severe degradation of signal coherence, thus undermining the overall performance of localization algorithms \cite{hardware_impairments, mimo_transmission, toward_fine_grained}.

Over the past decades, extensive research efforts have been dedicated to addressing phase synchronization challenges in multi-antenna and multi-carrier systems. Existing studies laid foundational techniques for frequency synchronization, notably Moose's pioneering cyclic prefix-based frequency offset estimation method, widely applied to mitigate \gls{cfo} in \gls{ofdm} systems \cite{technique_ofdm_frequency}. Subsequent advances introduced maximum likelihood estimators and least squares approaches, enhancing the accuracy and robustness of \gls{cfo} compensation techniques \cite{training_based_mimo}. More recently, studies have increasingly recognized the challenges introduced by spatial offsets inherent in massive \gls{mimo} architectures. Radhakrishnan et al.~\cite{hardware_impairments} and Studer et al.~\cite{mimo_transmission} highlighted significant localization performance degradation due to residual hardware impairments, particularly antenna-dependent phase variations. Shahmansoori et al.~\cite{position_orientation} emphasized the need for joint spatial-frequency synchronization strategies, integrating antenna calibration methods with channel estimation algorithms to improve localization precision. Despite these advancements, a comprehensive analysis of phase incoherence effects on localization performance with experimental validation remains limited, representing the research gap that this paper aims to address.

In this paper, we present a unified analytical and empirical study of phase incoherence in \gls{ofdm}-based massive \gls{mimo} localization, using two complementary tools: \acrfull{crlb} and \acrfull{saf}. We derive analytical expressions of the \gls{crlb} to reveal the mathematical optimal localization accuracy that the unbiased location estimator can achieve. Considering the \gls{crlb} does not take into account the global ambiguities that the phase offsets may cause, we develop a unified theoretical model based on the \gls{saf} and focus on sidelobe levels by defining a metric named \gls{pmsr}. To empirically validate our theoretical insights, we perform experimental measurements using a massive \gls{mimo} testbed and introduce a robust \gls{csi} calibration framework to compensate for frequency and spatial phase offsets in measured data. Experimental results demonstrate that our proposed calibration significantly improves the combination of coherent signals, thus greatly enhancing the localization performance.


The main contributions are summarized as follows:
\begin{itemize}
    \item We introduce an analytical framework that leverages the \gls{crlb} and \gls{saf}, explicitly accounting for both frequency and spatial phase offsets in \gls{ofdm}-based massive \gls{mimo} systems, thus providing theoretical insights into phase incoherence effects on localization performance.
    \item We systematically simulate and analyze the effects of phase incoherence on localization performance through introducing the \gls{pmsr} as a practical metric to quantify the ambiguity observed in \glspl{saf} under varying levels of phase incoherence.
    \item We propose a practical and effective \gls{csi} calibration framework that compensates for frequency and spatial phase offsets, enabling coherent signal combination across subcarriers and antennas.
    \item We validate the effectiveness of the proposed calibration approach through experimental measurements conducted on a real-world massive \gls{mimo} testbed, demonstrating a significant reduction of the localization \gls{rmse} (from meters to centimeter level) compared to uncalibrated scenarios.
\end{itemize}

\textbf{Scope and assumptions}: Throughout this paper, we assume a static single \gls{ue} and \gls{los} scenario. Multi-\gls{ue} is not considered due to the limited range resolution (i.e., \SI{16.67}{m}) of the validation testbed (Section~\ref{sec: real_world_validation}). Although the \gls{crlb} and \gls{saf} derivations assume only the LoS component for simplification, the observations remain valid in diffuse multipath where the non-\gls{los} terms are zero-mean with independent phases. We have preliminary evidence of the generalizability of the proposed CSI calibration framework in a dynamic passive sensing scenario. A complete evaluation is outside this paper’s scope and will be reported in our future work.

The remainder of this paper is structured as follows. Section~\ref{sec: system_model} details the system model and the corresponding \gls{crlb} and \gls{saf}. Section~\ref{sec: CRLB_simulation_results} evaluates the \gls{crlb} under varying phase offsets. Section~\ref{sec: SAF_simulation_results} analyzes \gls{saf} simulations with a focus on \gls{pmsr} trends. Section~\ref{sec: real_world_validation} provides experimental validation using a massive \gls{mimo} testbed in the real world. Finally, Section~\ref{sec: conclusion} summarizes the key findings of the paper and discusses directions for future research.

\section{System Model}
\label{sec: system_model}

\subsection{\gls{ofdm}-Based Massive \gls{mimo} System Model}

\begin{figure}[t]
\centerline{\includegraphics[width=0.8\columnwidth]{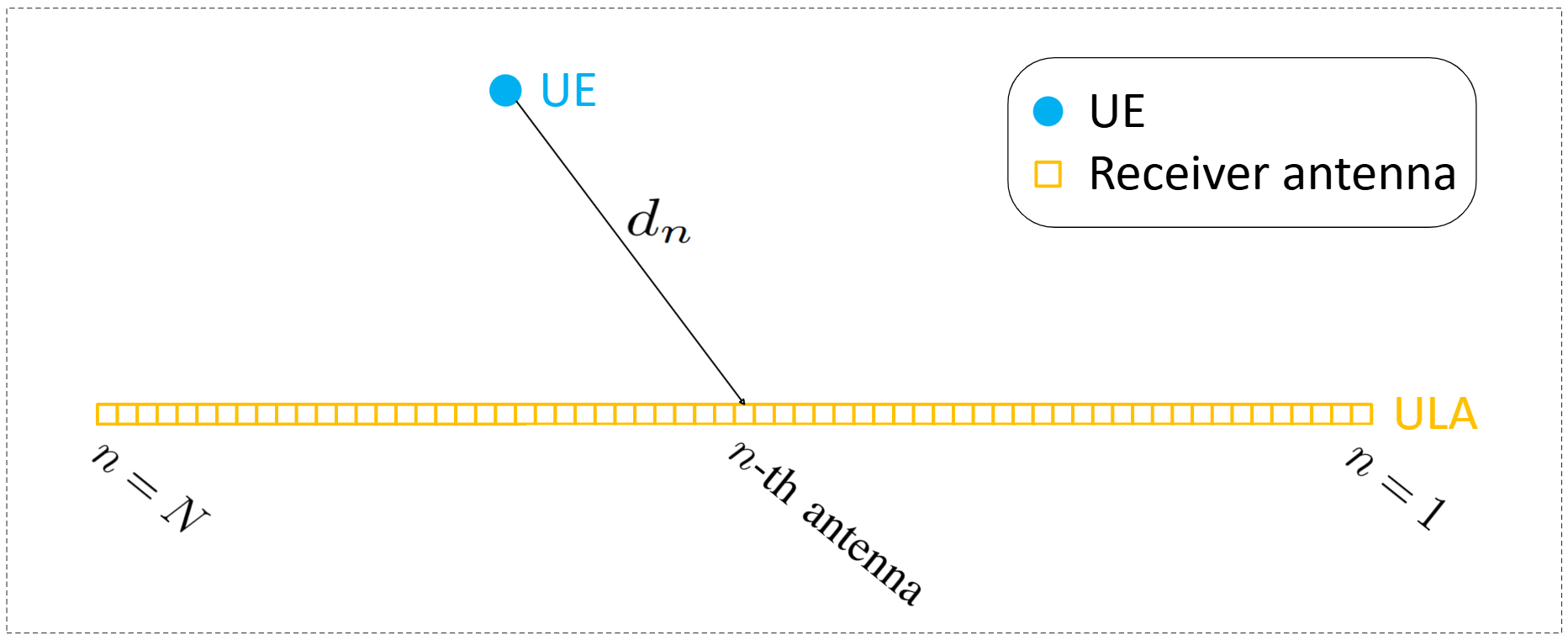}}
\caption{Considered scenario: A static \gls{ue} transmitting \gls{ofdm} signals with $K$ subcarriers to a \gls{ula} with $N$ receiver antennas.}
\label{fig: considered_scenario}
\end{figure}

To analyze the impact of phase incoherence on localization performance, we consider the \gls{ofdm}-based massive \gls{mimo} scenario depicted in Fig.~\ref{fig: considered_scenario}. A static \acrfull{ue} transmits a predefined pilot signal to $N$ receiver antennas arranged as a \gls{ula}. The signal received at each antenna differs according to its respective distance $d_n$ from the \gls{ue}. The received signal at the $n$-th antenna during a single \gls{ofdm} symbol is expressed as:
\begin{equation} 
    r_{nk}=\alpha _{n}e^{-j2\pi \left({\frac {f_c}{c}+\frac {\Delta f}{c}k}\right)d_{n}} + w_{nk},
    \label{eq: ideally_received_signal}
\end{equation} 
where $k \in \left\{1,2,\dots,K\right\}$ is the \gls{ofdm} subcarrier index, $\alpha_{n}$ denotes the signal amplitude accounting for path loss, $f_c$ is the carrier frequency, $\Delta f$ is the subcarrier spacing, and $c$ is the speed of light. $w_{nk}$ is the complex additive white Gaussian noise, and $w_{nk} \sim \mathcal{CN}(0, \sigma_n^2)$.

\begin{figure}[t]
    \centering
    \begin{subfigure}[b]{0.49\linewidth}
        \centering
        \includegraphics[width=\linewidth]{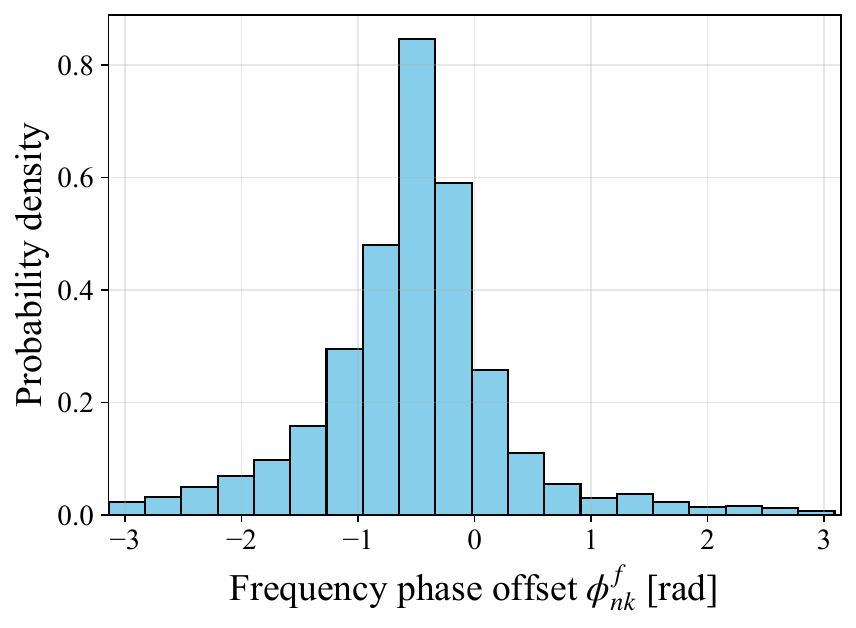} 
        \caption{}
    \end{subfigure}
    \begin{subfigure}[b]{0.49\linewidth}
        \centering
        \includegraphics[width=\linewidth]{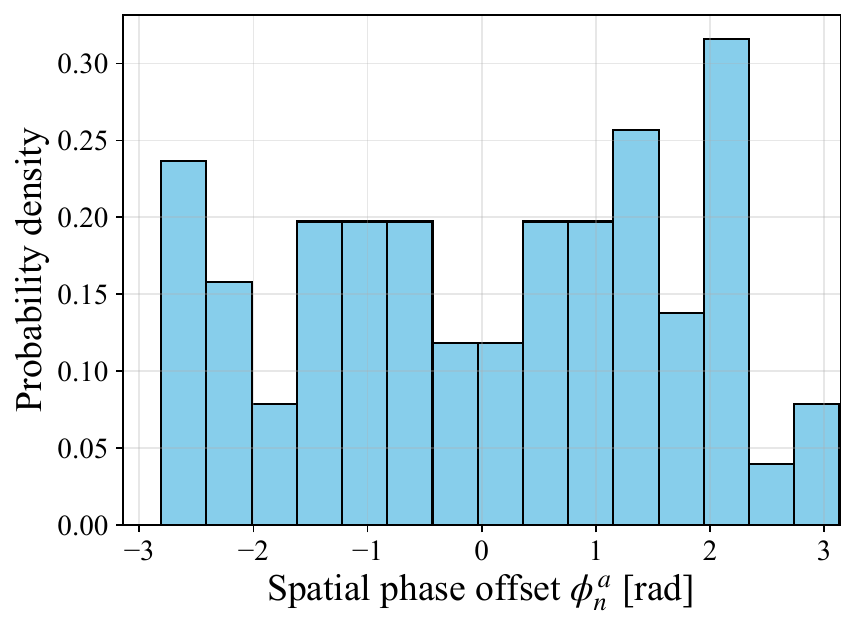}  
        \caption{}
    \end{subfigure}

    \caption{Distributions of the frequency and spatial phase offsets from the measured data: (a) Frequency phase offset: Gaussian distribution. (b) Spatial phase offset: uniform distribution.}
    \label{fig: distributions_offsets}
\end{figure}

When frequency-dependent phase offset is present, an additional phase term $\phi_{nk}^{f}$ is introduced. 
This offset is commonly modelled as a Gaussian distributed random variable: $\phi_{nk}^{f} \sim \mathcal{N} (\mu_{\phi}, \sigma_{\phi}^{f2})$ \cite{OFDM_systems_presence}. As shown in Fig.~\ref{fig: distributions_offsets}(a), this Gaussian distribution is empirically validated using the data measured in Section~\ref{sec: real_world_validation}. Similarly, an antenna-dependent phase offset $\phi_{n}^{a}$ is modelled as a uniform distribution: $\phi_n^a \sim \mathcal{U}(-\Delta,\Delta)$, with a standard deviation $\sigma_{\phi}^a$, as shown in Fig.~\ref{fig: distributions_offsets}(b). Consequently, the received signal in (\ref{eq: ideally_received_signal}) becomes:
\begin{equation}
    \tilde r_{nk} = \alpha_n e^{-j2\pi\left(\frac {f_c}{c} + \frac{\Delta f}{c}k\right) d_n} e^{j\phi_{nk}^f} e^{j\phi_{n}^a} + w_{nk}.
\end{equation}

\subsection{\acrlong{crlb}}
\label{subsec: CRLB}
We respectively analyze the impacts of frequency and spatial phase offsets by deriving their corresponding analytical expressions of \gls{crlb} for estimating the \gls{ue} location.

Define the unknown parameter vector as: 
\begin{equation}
    \boldsymbol{\eta} = \begin{bmatrix} x & y & \boldsymbol{\phi} \end{bmatrix}^T,
\end{equation}
where $(x,y)$ is the \gls{ue} location, $\boldsymbol{\phi}=\left\{ \phi_{nk}^f \right\}$ or $\left\{ \phi_{n}^a \right\}$, depending on the considered phase offset. Then the full \gls{fim} with prior knowledge of $\boldsymbol{\phi}$ is structured as \cite{fundamental_limits_optimization}:
\begin{equation}
    \mathbf{J}_\eta =
    \begin{bmatrix}
    \mathbf{J}_{xy,xy} & \mathbf{J}_{xy,\phi} \\
    \mathbf{J}_{\phi,xy} & \mathbf{J}_{\phi,\phi}
    \end{bmatrix}.
\end{equation}
To marginalize the effect of $\boldsymbol{\phi}$, an \gls{efim} for $\mathbf{p} = [x, y]^T$ is given by:
\begin{equation}
    \mathbf{J}_{\text{eff}} = \mathbf{J}_{xy,xy} - \mathbf{J}_{xy,\phi} \cdot \mathbf{J}_{\phi,\phi}^{-1} \cdot \mathbf{J}_{xy,\phi}^T,
\end{equation}
where:
\begin{equation}
    \mathbf{J}_{xy,xy} = \frac{2}{\sigma_n^2} \sum_{n=1}^{N} \sum_{k=1}^{K}
    \Re\left\{
    \frac{\partial \mu_{nk}}{\partial \mathbf{p}}
    \left( \frac{\partial \mu_{nk}}{\partial \mathbf{p}} \right)^H
    \right\};
\end{equation}
\vspace*{0.05in}
\begin{equation}
    \mathbf{J}_{xy,\phi} = \frac{2}{\sigma_n^2} \sum_{n=1}^{N} \sum_{k=1}^{K}
    \Re\left\{
    \frac{\partial \mu_{nk}}{\partial \mathbf{p}}
    \left( \frac{\partial \mu_{nk}}{\partial \boldsymbol{\phi}} \right)^H
    \right\},
\end{equation}
and $\Re \left\{ \cdot \right\}$ denotes the real part of a complex number. $\mu_{nk}$ is the expected value of the received signal, $\mu_{nk} = \mathbb{E}(\tilde r_{nk})$, with:
\begin{align}
    \frac{\partial \mu_{nk}}{\partial x} &= -j2\pi\left(\frac{f_c}{c} + \frac{\Delta f}{c}k \right) \frac{x - x_n}{d_n} \mu_{nk}; \\
    \frac{\partial \mu_{nk}}{\partial y} &= -j2\pi\left(\frac{f_c}{c} + \frac{\Delta f}{c}k \right) \frac{y - y_n}{d_n} \mu_{nk}; \\
    \frac{\partial \mu_{nk}}{\partial \phi_{nk}^f} &= j \mu_{nk} \quad \text{or} \quad \frac{\partial \mu_{nk}}{\partial \phi_{n}^a} = j \mu_{nk}.
\end{align}
Furthermore, in the case of $\phi_{nk}^f$ with the Gaussian distribution:
\begin{equation}
    \mathbf{J}_{\phi,\phi} = \frac{2}{\sigma_n^2} \operatorname{diag}\left( |\mu_{nk}|^2 \right) + \frac{1}{\sigma_\phi^2} \mathbf{I}_{NK};
\end{equation}
and for $\phi_{n}^a$ with the uniform distribution:
\begin{equation}
    \mathbf{J}_{\phi,\phi} = \frac{2}{\sigma_n^2} \operatorname{diag}\left( \sum_{k=1}^{K} |\mu_{nk}|^2 \right) + \frac{3}{\Delta^2} \mathbf{I}_N.
\end{equation}
Finally, we evaluate the \acrfull{rmse} of the location estimation as:
\begin{equation}
    \text{CRLB}_{xy}^{\text{RMSE}} = \sqrt{\text{tr}(\text{CRLB}_{xy})} = \sqrt{\text{tr}(\mathbf{J}_{\text{eff}}^{-1})}
    \label{eq: crlb_rmse}
\end{equation}
where the operator $\text{tr}(\cdot)$ denotes the sum of the diagonal terms in $\text{CRLB}_{xy}$.


\subsection{Spatial Ambiguity Function and Performance Metric}
\label{subsec: SAF}
We compare the \glspl{saf} in the three cases: ideal signal, signal with frequency phase offset, and signal with spatial phase offset. Using the ideally received signal defined in (\ref{eq: ideally_received_signal}), the \gls{saf} is formulated as the matched filter output that visualizes the spatial correlation between the true signals and the expected signals from all test locations $(x,y)$ \cite{near_field_coherent_radar, wavefield_networked_sensing}. The \gls{saf} is expressed as:
\begin{equation}
\mathcal{A}(x,y)=\left|\sum_{n=1}^{N}\sum_{k=1}^{K}\mathcal{A}_{nk}(x,y)\right|^{2},
\label{eq: SAF}
\end{equation}
where
\begin{equation}
\mathcal{A}_{nk}(x,y)=e^{j2\pi\left(\frac{f_c}{c}+\frac{\Delta f}{c}k\right)\left(d_n(x,y)-d_n\right)}.
\label{eq: AF_ideal}
\end{equation}
When frequency phase offset $\phi_{nk}^{f}$ is present, the ambiguity function is expressed as:
\begin{equation}
    \mathcal {A}_{nk}^{f}(x,y) =e^{j2\pi \left ({\frac {f_c}{c}+\frac {\Delta f}{c}k}\right) \left(d_n(x,y)-d_n \right)} \cdot e^{j \phi_{nk}^{f}}.
    \label{eq: AF_frequency_offset}
\end{equation}
Similarly, the ambiguity function incorporating spatial phase offset $\phi_{n}^{a}$ is expressed as:
\begin{equation}
    \mathcal {A}_{nk}^{a}(x,y) =e^{j2\pi \left ({\frac {f_c}{c}+\frac {\Delta f}{c}k}\right) \left(d_n(x,y)-d_n \right)} \cdot e^{j \phi_{n}^{a}}.
    \label{eq: AF_channel_offset}
\end{equation}

We aim to evaluate the ambiguities caused by the phase offsets, which can be revealed by the sidelobe levels in the \gls{saf}. Therefore, we define a performance metric named \gls{pmsr}, which is the ratio of the \gls{saf} peak power to the median sidelobe power:
\begin{equation}
    \mathrm {PMSR} = \frac { \lvert \mathcal A \left (x_\text{sen}, y_\text{sen}\right ) \rvert ^{2} }{ \operatorname{median} \limits _{ (x,y)\in \mathcal {R}\setminus \Omega } \lvert \mathcal{A} \left ({{x,y}}\right ) \rvert ^{2}},
\end{equation}
where the \gls{saf} peak corresponds to the sensed \gls{ue} location $\left (x_\text{sen}, y_\text{sen}\right )$. $\mathcal {R}$ denotes the full spatial region for evaluating the \gls{saf}, and $\Omega \in \mathcal{R}$ represents the main lobe region. \gls{pmsr} effectively characterizes the level of the sidelobes in the \gls{saf}. The higher \gls{pmsr} values indicate reduced sidelobe interference, thereby enabling a more accurate localization.


\section{Cramér-Rao Lower Bound Simulation Results}
\label{sec: CRLB_simulation_results}

In this section, we analyze the $\text{CRLB}_{xy}^\text{RMSE}$ derived in Section~\ref{subsec: CRLB}, under varying levels of frequency and spatial phase offsets. To ensure a meaningful comparison with further real-world validation results, the simulation parameters are aligned with the practical massive \gls{mimo} testbed, as described in Section~\ref{sec: real_world_validation}. The complete list of parameters is summarized in Table~\ref{tab: mamimo_param}.

\subsection{Impact of Frequency Phase Offset}
\label{subsec: crlb_result_frequency_offset}
Fig.~\ref{fig: crlb}(a) shows the $\text{CRLB}_{xy}^{\text{RMSE}}$ as a function of the standard deviation $\sigma_{\phi}^f$ of the Gaussian-distributed frequency phase offset with different antenna numbers. As $\sigma_{\phi}^f$ increases from $\frac{\pi}{4}$ to $\pi$, the $\text{CRLB}_{xy}^{\text{RMSE}}$ increases by a factor of 5 to 15 compared to the ideal $\text{CRLB}_{xy}^{\text{RMSE}}$. This observation indicates that frequency phase distortions, even with large variances, induce only minor degradation in localization accuracy. The robustness against frequency phase offsets arises because such distortions are subcarrier-specific and tend to be averaged out across the multiple subcarriers. 

On the other hand, increasing the number of antennas significantly reduces the $\text{CRLB}_{xy}^{\text{RMSE}}$ due to the enhanced spatial diversity. For example, when a localization accuracy of \SI{1}{cm} is desired under the presence of $\sigma_{\phi}^f=\pi$, the frequency phase offset does not need to be resolved when using more than 23 antennas. In other words, a preferred localization accuracy can be achieved by either eliminating the frequency phase distortions or using a larger antenna array.

\subsection{Impact of Spatial Phase Offset}
\label{subsec: crlb_spatial_offset}
\begin{figure}[tbp]
    \centering
    \begin{subfigure}[b]{0.49\linewidth}
        \centering
        \includegraphics[width=\linewidth]{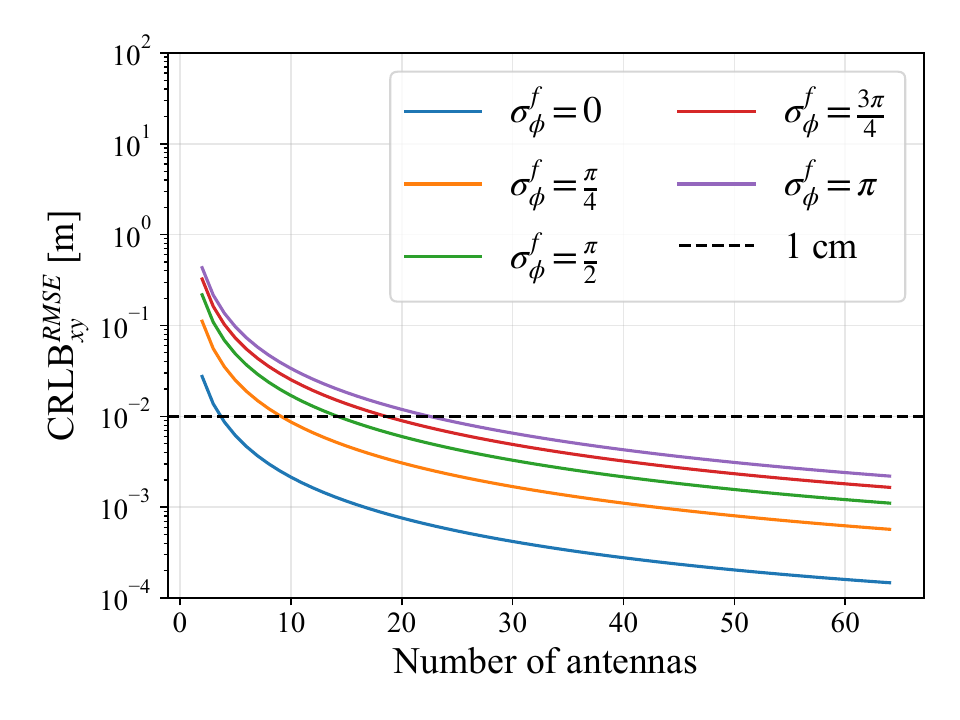} 
        \caption{}
    \end{subfigure}
    \begin{subfigure}[b]{0.49\linewidth}
        \centering
        \includegraphics[width=\linewidth]{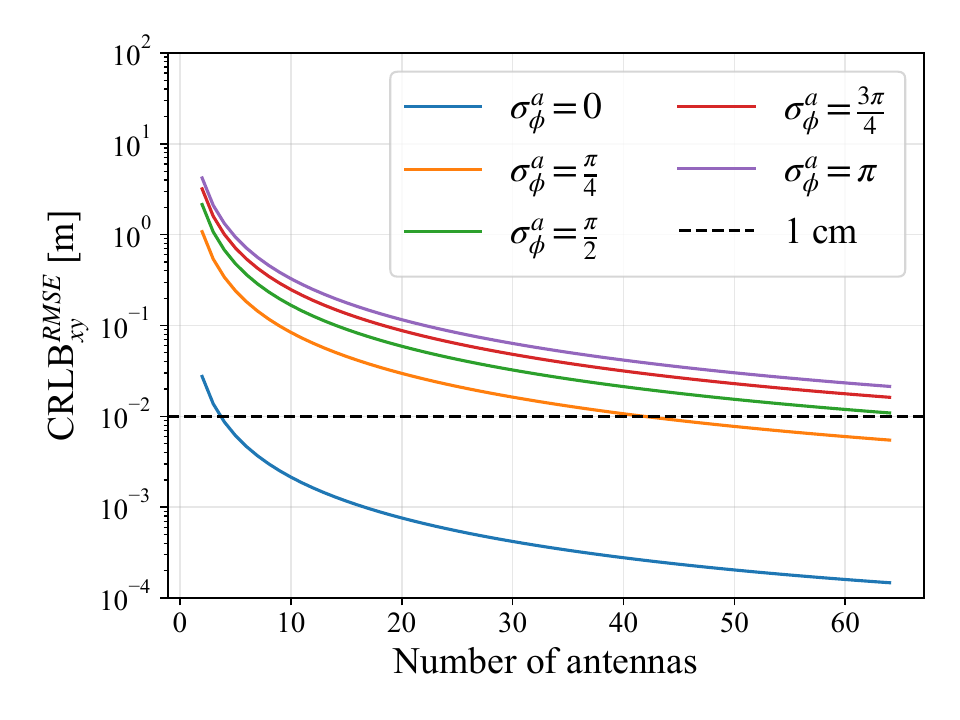}  
        \caption{}
    \end{subfigure}

    \caption{\gls{crlb}-based localization \gls{rmse} versus phase incoherence: (a) Frequency phase offset: minor impact on localization accuracy. (b) Spatial phase offset: severe degradation of localization accuracy.}
    \label{fig: crlb}
\end{figure}

Fig.~\ref{fig: crlb}(b) illustrates the $\text{CRLB}_{xy}^{\text{RMSE}}$ as a function of the standard deviation $\sigma_\phi^a$ of the uniformly distributed spatial phase offset. A remarkable increase in the localization error is observed as $\sigma_\phi^a$ grows. The growth of $\text{CRLB}_{xy}^{\text{RMSE}}$ is 30 to 130 times compared to the ideal $\text{CRLB}_{xy}^{\text{RMSE}}$. This behavior highlights the severe sensitivity of localization performance to spatial phase incoherence. In contrast to frequency phase offset, spatial phase offset directly disrupts the constructive combining of signals across the antenna array.

Moreover, although the $\text{CRLB}_{xy}^{\text{RMSE}}$ also degrades significantly as the number of antennas increases in this case, a localization accuracy of \SI{1}{cm} is hard to achieve with a reasonable number of antennas when the spatial phase distortions exist.


\section{Spatial Ambiguity Function Simulation Results}
\label{sec: SAF_simulation_results}

This section presents simulation results of the \glspl{saf} and \gls{pmsr} defined in Section~\ref{subsec: SAF}. The simulation parameters follow the practical testbed in Table~\ref{tab: mamimo_param} as well.

\subsection{Spatial Ambiguity Function of Ideal Signal}
Fig.~\ref{fig: SAF_ideal_data} presents the \gls{saf} computed from the ideal signal model with \gls{ue} location being $(x_{\text{true}}, y_{\text{true}})=(-2, 1)$. The x- and y-cuts of the \gls{saf} are also presented to indicate the sidelobe levels clearly. The \gls{saf} exhibits a sharp mainlobe and well-suppressed sidelobes. The high \gls{pmsr} of 29.45 dB reflects 
minimal ambiguity.

\begin{figure}[tbp]
    \centering
    \begin{subfigure}[b]{0.9\linewidth}
        \centering
        \includegraphics[width=\linewidth]{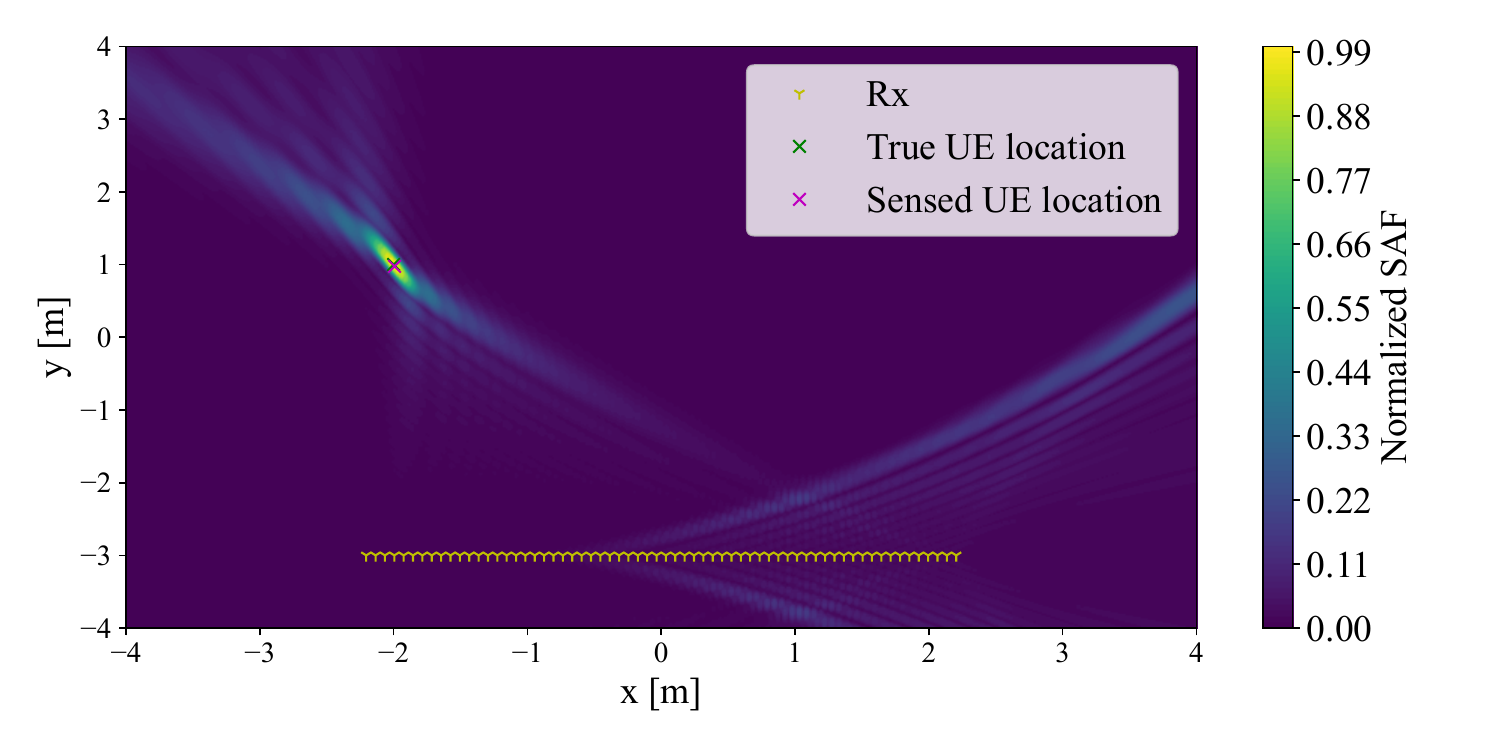}
        \caption{}
        \label{fig: SAF_ideal_data}
    \end{subfigure}
    
    
    \begin{subfigure}[b]{\linewidth}
        \centering
        \includegraphics[width=\linewidth]{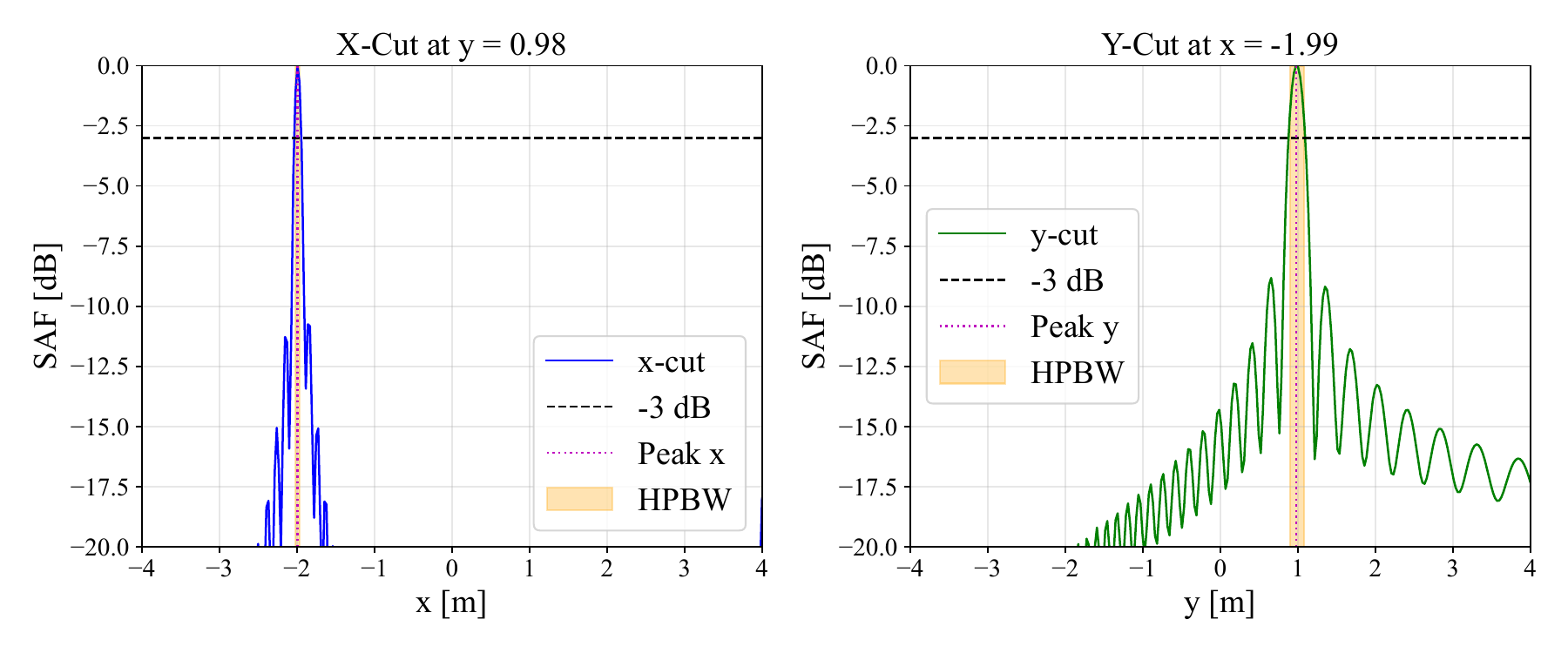}  
        \caption{}
        \label{fig: SAF_XY_Cut_ideal_data}
    \end{subfigure}

    \caption{\gls{saf} of the simulated ideal signal with $(x_{\text{true}}, y_{\text{true}})=(-2, 1)$: (a) \gls{saf}; (b) X- and Y-cuts of the \gls{saf} at the sensed \gls{ue} location. \gls{pmsr} = \SI{29.45}{dB}.}
    \label{fig: SAF_ideal_data}
\end{figure}

\subsection{Spatial Ambiguity Function with Frequency Phase Offset}
\label{subsec: SAF_frequency_offset}

\begin{figure}[t]
    \centering
    \begin{subfigure}[b]{0.9\linewidth}
        \centering
        \includegraphics[width=\linewidth]{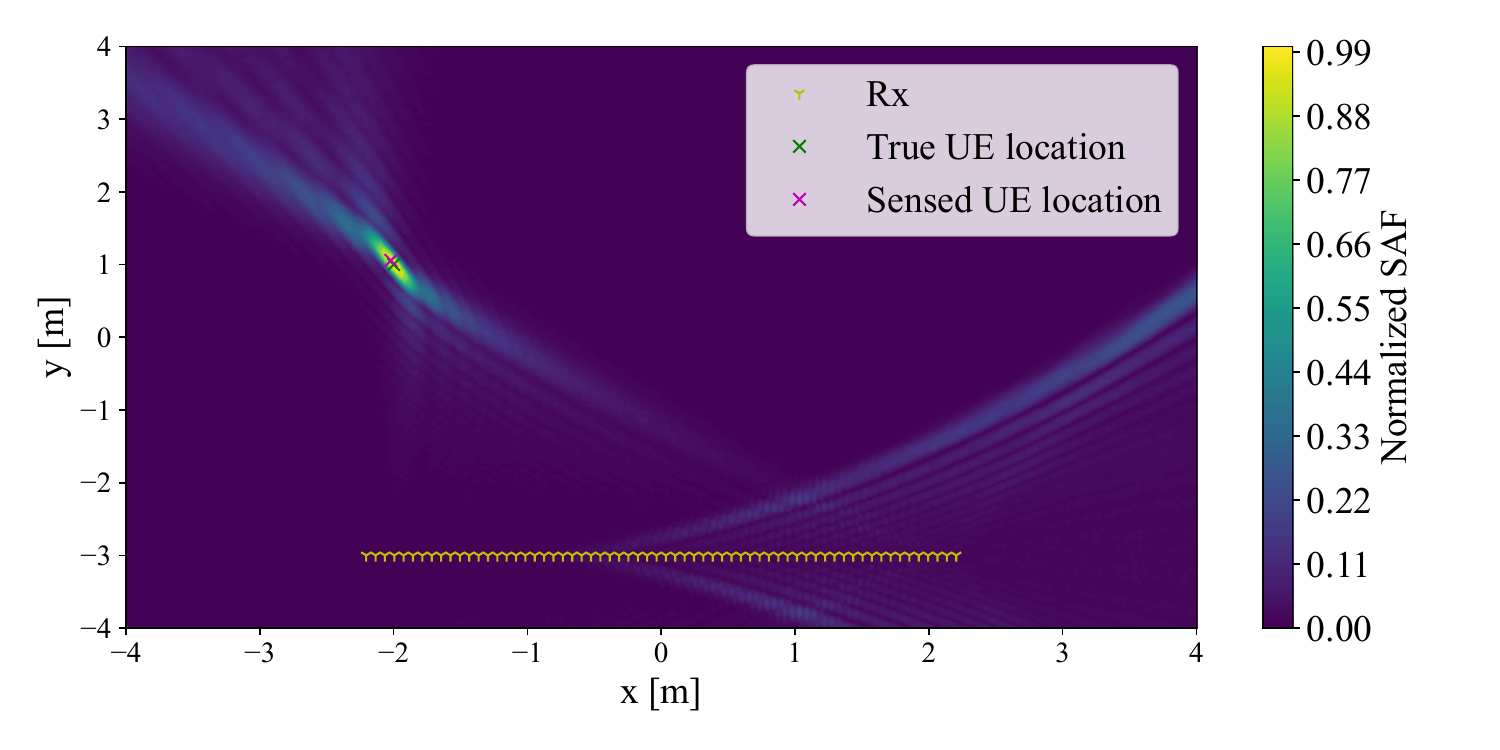}
        \caption{}
    \end{subfigure}
    
    
    \begin{subfigure}[b]{\linewidth}
        \centering
        \includegraphics[width=\linewidth]{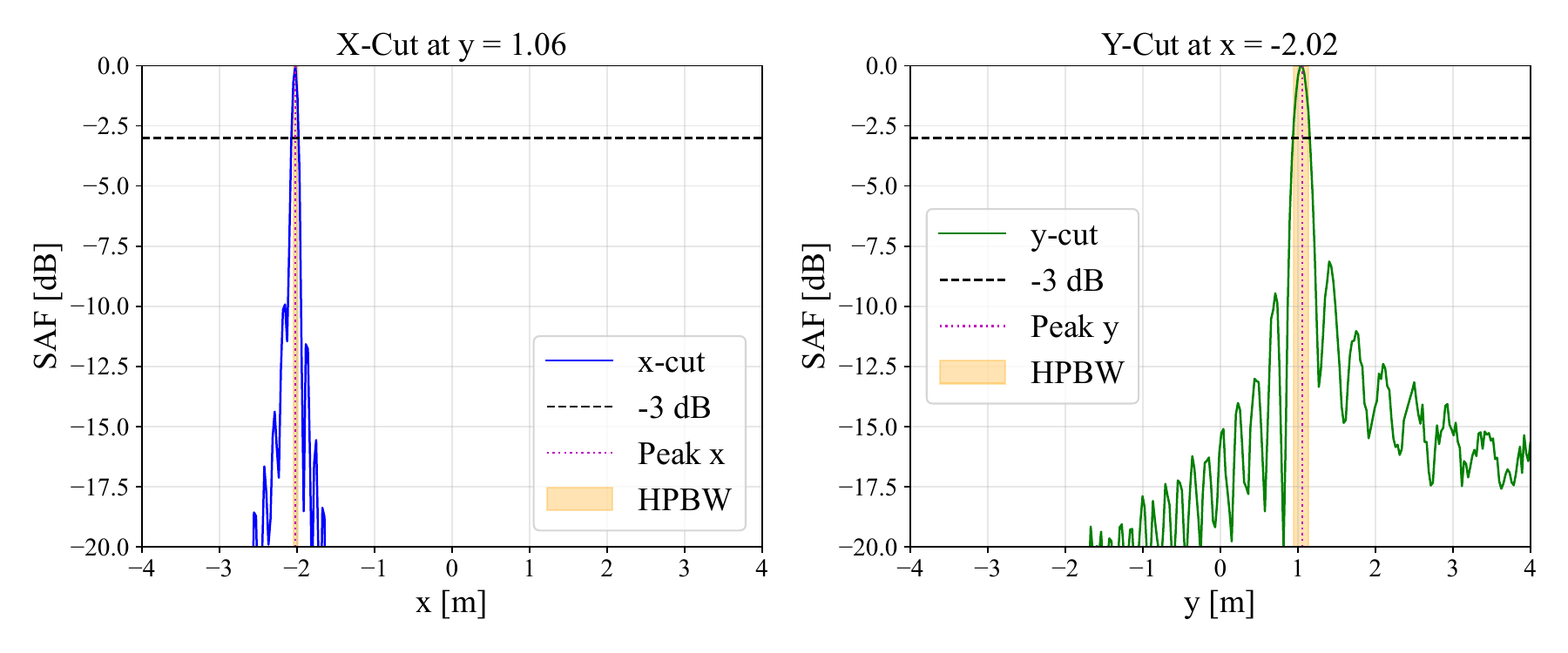}
        \caption{}
    \end{subfigure}

    \caption{\gls{saf} of the simulated signal with frequency phase offset ($\sigma_\phi^f=\frac{\pi}{4}$): (a) \gls{saf}; (b) X- and Y-cuts of the \gls{saf} at the sensed \gls{ue} location. \gls{pmsr} = \SI{28.69}{dB}.}
    \label{fig: SAF_frequency_offset}
\end{figure}

To assess the influence of frequency phase offset, we simulate the \gls{saf} under varying standard deviation values of the Gaussian-distributed offset. Specifically, we set $\sigma_{\phi}^f = \left \{\frac{\pi}{16}, \frac{\pi}{8},\frac{3\pi}{16},\frac{\pi}{4},\frac{5\pi}{16},\frac{3\pi}{8},\frac{7\pi}{16},\frac{\pi}{2} \right \}$. Fig.~\ref{fig: SAF_frequency_offset} shows one of the resulting \glspl{saf} with $\sigma_{\phi}^f = \frac{\pi}{4}$. The \gls{saf} results reveal that the frequency phase offset has a minor visual impact on the \gls{saf} structure. This observation remains consistent with the other values of $\sigma_{\phi}^f$. The corresponding trend of \gls{pmsr}, as plotted in Fig.~\ref{fig: PMSR} (blue line), further confirms that the frequency phase offset results in mild variation in localization performance, consistent with \gls{crlb} results in Section~\ref{subsec: crlb_result_frequency_offset}. This is because the impact tends to average out in matched filtering across the subcarrier dimension.

\subsection{Spatial Ambiguity Function with Spatial Phase Offset}
\label{subsec: SAF_spatial_offset}

\begin{figure}[ht]
    \centering
    \begin{subfigure}[b]{0.9\linewidth}
        \centering
        \includegraphics[width=\linewidth]{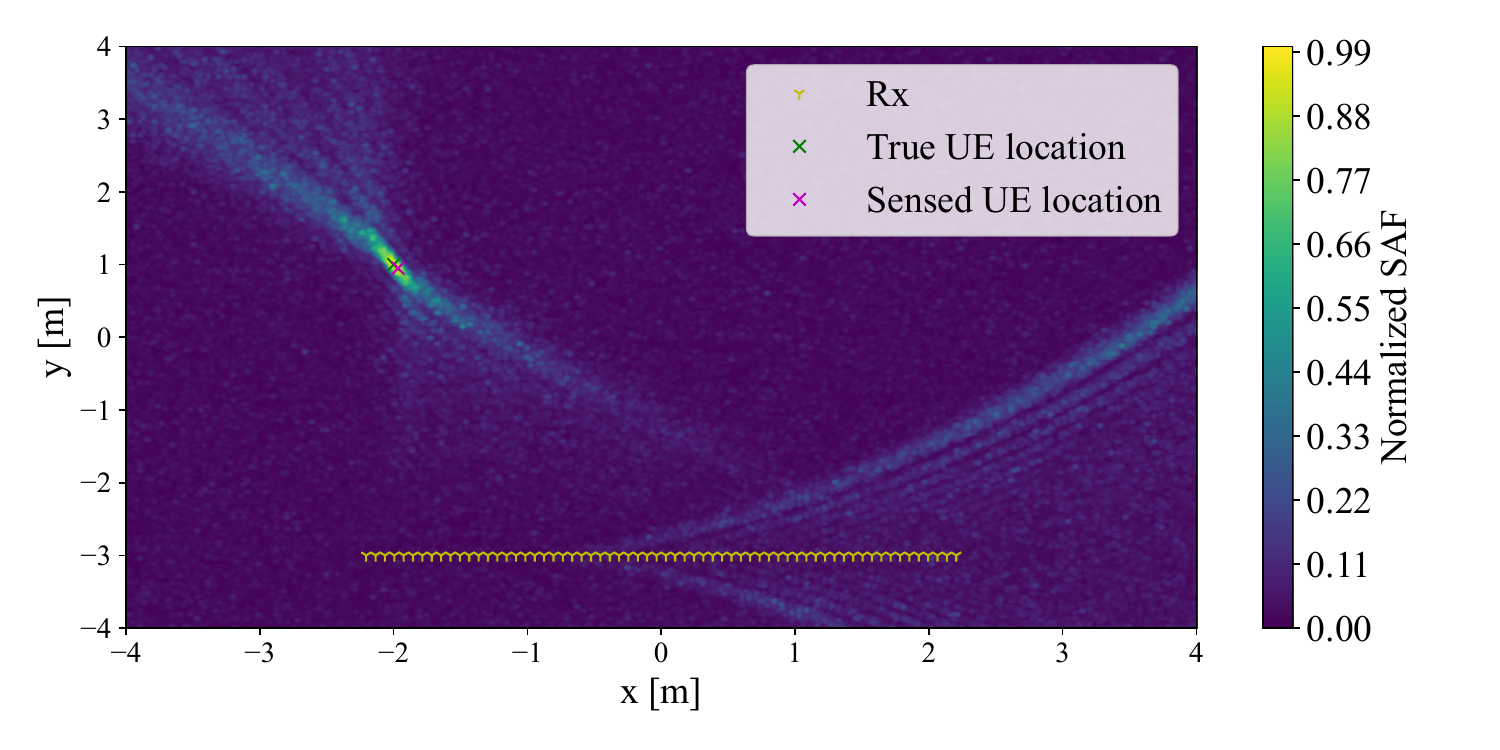}
        \caption{}
    \end{subfigure}
    
    
    \begin{subfigure}[b]{\linewidth}
        \centering
        \includegraphics[width=\linewidth]{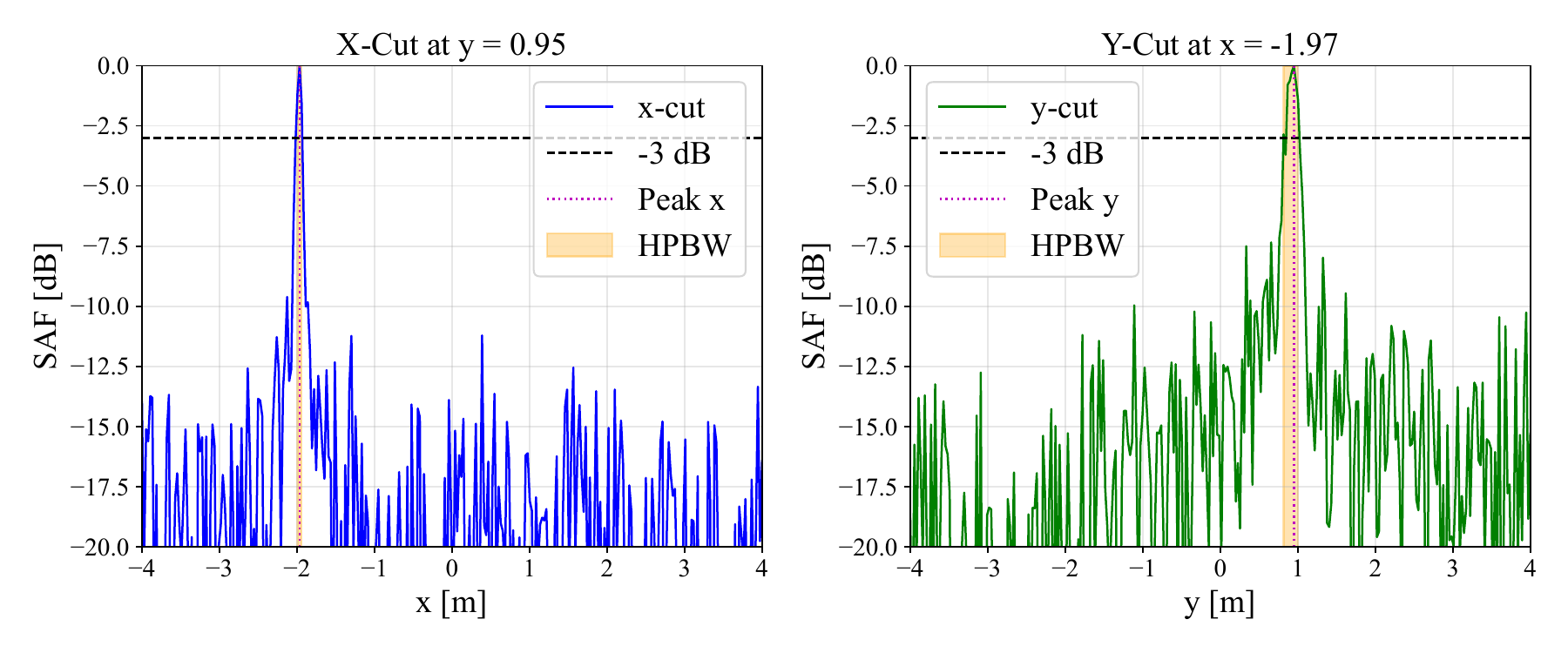}
        \caption{}
    \end{subfigure}

    \caption{\gls{saf} of the simulated signal with spatial phase offset ($\sigma_\phi^a=\frac{\pi}{4}$): (a) \gls{saf}; (b) X- and Y-cuts of the \gls{saf} at the sensed \gls{ue} location. \gls{pmsr} = \SI{18.48}{dB}.}
    \label{fig: SAF_channel_offset}
\end{figure}

\begin{figure}[htbp]
\centerline{\includegraphics[width=0.8\columnwidth]{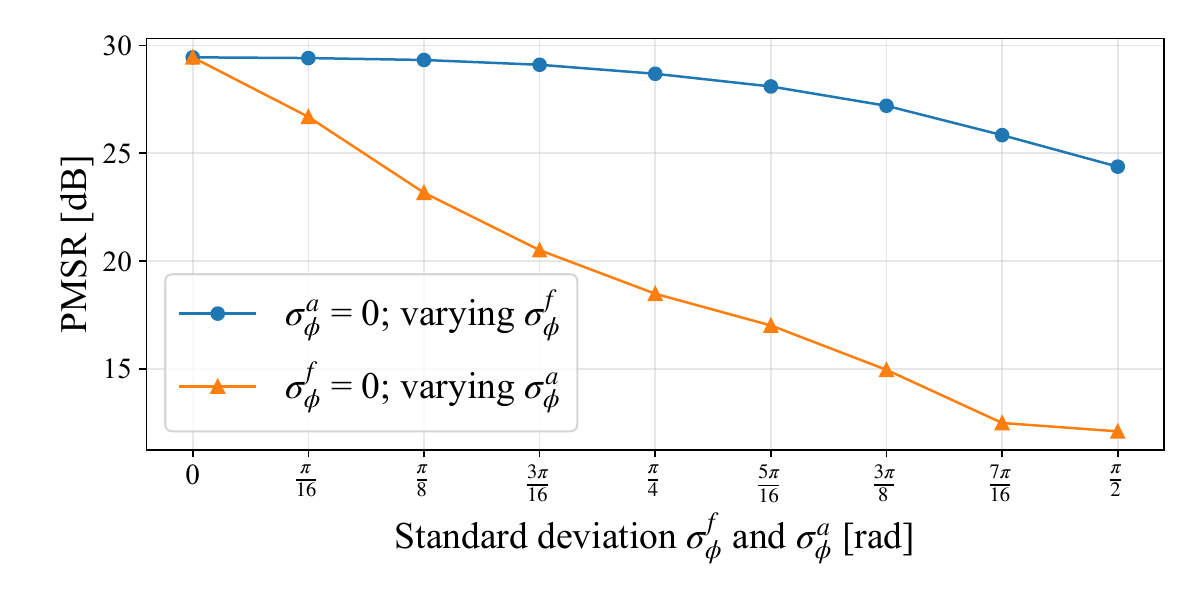}}
\caption{\gls{pmsr} degradation with increasing frequency phase offset (blue line) and spatial phase offset (orange line).}
\label{fig: PMSR}
\end{figure}

Similarly to the frequency phase offset case, we simulate the \gls{saf} under varying levels of spatial phase offsets. Fig.~\ref{fig: SAF_channel_offset} presents the resulting \gls{saf} with $\sigma_\phi^a = \frac{\pi}{4}$. Pronounced sidelobe growth is evident, reducing \gls{pmsr} to \SI{18.48}{dB}. Unlike frequency phase offset, spatial phase mismatches persist across antennas and directly disrupt coherent signal combination. Fig.~\ref{fig: PMSR} (orange line) shows a significant \gls{pmsr} drop as $\sigma_\phi^a$ increases, mirroring the sharp \gls{crlb} degradation seen in Section~\ref{subsec: crlb_spatial_offset}.

\section{Real-World Measurement Validation}
\label{sec: real_world_validation}
In this section, we apply real-world measurements to validate our theoretical findings. The measured \gls{csi} data is processed from a calibration perspective, where frequency and spatial phase offsets are sequentially estimated and compensated using extrinsic calibration with a known \gls{ue} location. 

\subsection{Measurement Setup and Data Collection}
\begin{figure}[t]
\centerline{\includegraphics[width=0.8\columnwidth]{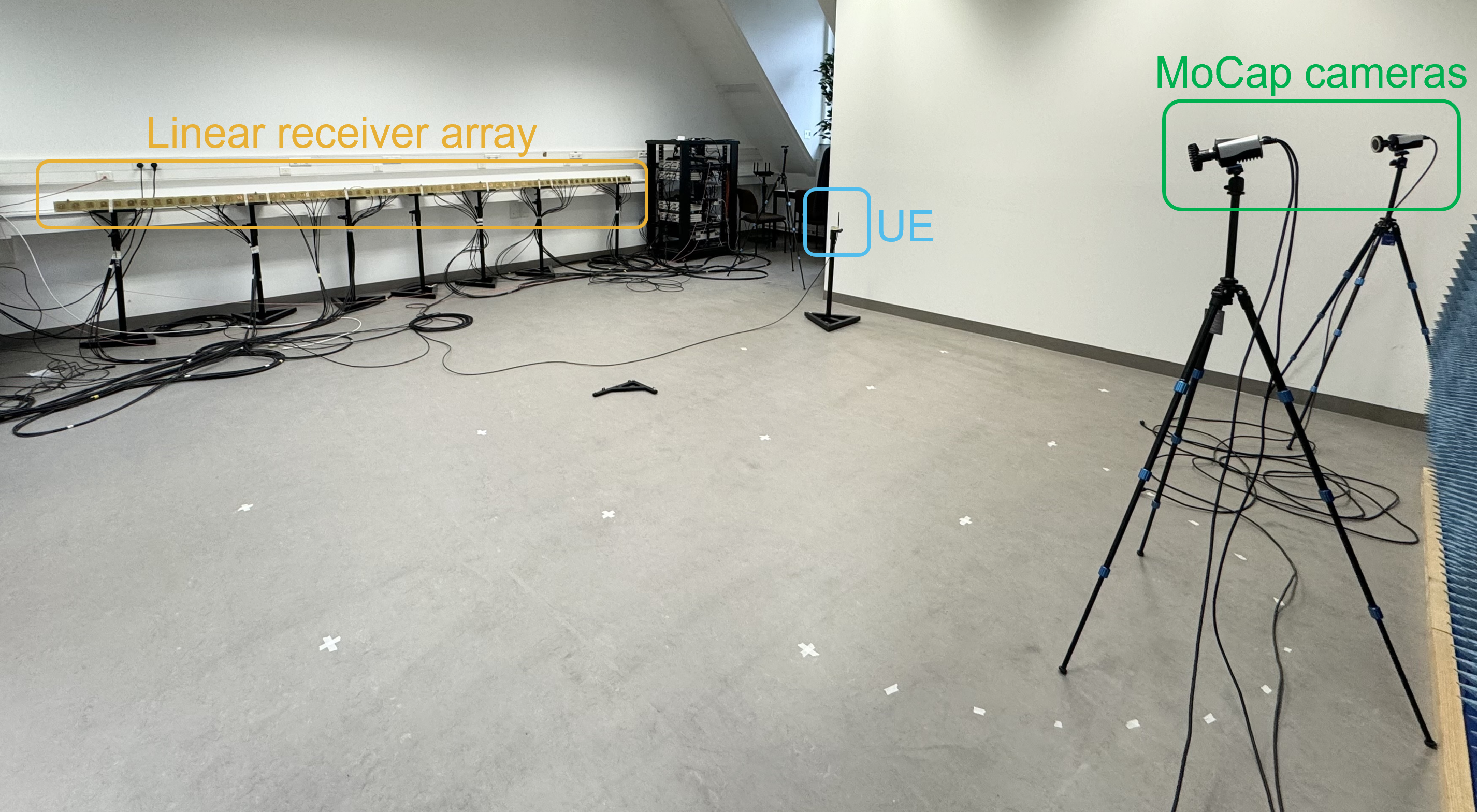}}
\caption{Measurement setup with massive \gls{mimo} testbed and motion capture cameras.}
\label{fig: measurement_scenario}
\end{figure}

 The practical measurement scenario is shown in Fig.~\ref{fig: measurement_scenario}. The setup comprises the massive \gls{mimo} testbed at KU Leuven and a Qualisys \gls{mocap} system~\cite{ni_mamimo, mocap_sheet}. The massive \gls{mimo} system operates in time division duplex mode, using \gls{ofdm} for signal transmission. The \gls{ue}, equipped with a single dipole antenna, transmits a predefined pilot signal. The base station is equipped with 64 patch antennas arranged as a linear array to receive the pilot signals and perform channel estimation. The detailed system parameters are listed in Table~\ref{tab: mamimo_param}. To obtain the ground truth location of the \gls{ue}, the \gls{mocap} system is calibrated within the same coordinate system as the \gls{mimo} testbed. It provides millimeter-level accuracy measurement by detecting a marker attached to the \gls{ue}. The \gls{csi} data is collected for \SI{5}{\second}, corresponding to $L = 10000$ \gls{ofdm} symbols. The resulting measured \gls{csi} is represented as $\textbf{r} \in \mathbb{C}^{N \times K \times L}$.

\begin{table}[t]
\caption{Massive \gls{mimo} system parameters}
\begin{center}
\begin{tabular}{|l|l|l|}
\hline
\textbf{System parameter} & \textbf{Symbol} & \textbf{Value}   \\ \hline
Center frequency           & $f_c$           & \SI{3.5}{\GHz}   \\ \hline
Bandwidth                  & BW              & \SI{18}{\MHz}    \\ \hline
Sampling rate              & $f_s$           & \SI{2}{\kHz}     \\ \hline
Number of subcarriers      & K               & 100              \\ \hline
Subcarrier spacing         & $\Delta f$      & \SI{180}{\kHz}   \\ \hline
Number of \gls{ue}               & M               & 1                \\ \hline
Number of receiver antennas& N               & 64                \\ \hline
Antenna spacing            & $\Delta a$      & \SI{0.07}{\m}     \\ \hline
\end{tabular}
\label{tab: mamimo_param}
\end{center}
\end{table}

\subsection{Phase Synchronization Algorithms}


\subsubsection{Frequency Phase Offset Estimation and Compensation}
First, the frequency phase offset is estimated using the \gls{lse}. Considering that the \gls{ue} remains static in a stationary environment, the phase offsets are assumed to be time-consistent. Therefore, with the ground truth location provided by the \gls{mocap} system, the ideal received signal $r_{nk}(l)$ can be modeled as in (\ref{eq: ideally_received_signal}). Then, the frequency phase offset can be estimated as: 
\begin{equation}
    \hat{\phi}_{nk}^f= \arg \min_{\phi_{nk}^f} \sum_{l=1}^{L} \left | \tilde r_{nk}(l)-e^{j\phi_{nk}^f} \cdot r_{nk}(l) \right |^2 ,
\end{equation}
where $\tilde r_{nk}(l)$ is the measured signal, and $l = {1,2,...,L}$ is the symbol index. The optimal solution of the \gls{lse} is given by:
\begin{equation}
    e^{j \hat \phi_{nk}^f} = \mathbf{ \frac{r_{nk}^H \cdot \tilde{r}_{nk}}{r_{nk}^H \cdot r_{nk}} },
    \label{eq: LSE}
\end{equation}
where $(\cdot)^H$ denotes the Hermitian transpose \cite{fundamentals}. As a result, the compensated signal can be simply derived as
\begin{equation}
    \hat r_{nk}(l) = \tilde{r}_{nk}(l) \cdot e^{-j \hat \phi_{nk}^f}.
\end{equation}

\subsubsection{Spatial Phase Offset Estimation and Compensation}
The spatial phase offset is derived and compensated for after performing the matched filter across the subcarrier dimension. Considering the single \gls{ue} scenario, the peak value along the range dimension is the \gls{los} signal from the \gls{ue}. As a result of the range processing, $N$ \gls{los} signals with spatial phase offset calibration are used to compute the localization image, where the phase offset $e^{j\hat \phi_n^a}$ is estimated by comparing the measured signal with the ideal signal $r_n^{LoS} = e^{-j2\pi \frac {f_c}{c} d_n}$ as:
\begin{equation}
     e^{j\hat \phi_n^a} = \frac{\tilde r_n^{LoS}}{r_n^{LoS}}.
\end{equation}
Then the compensated signal becomes:
\begin{equation}
    \hat r_n^{LoS} = \tilde r_n^{LoS} \cdot e^{-j \hat \phi_n^a}.
\end{equation}

\subsection{Validation Results}

\begin{figure*}[htbp]
    \centering
    \begin{subfigure}[b]{0.32\linewidth}
        \centering
        \includegraphics[width=\linewidth]{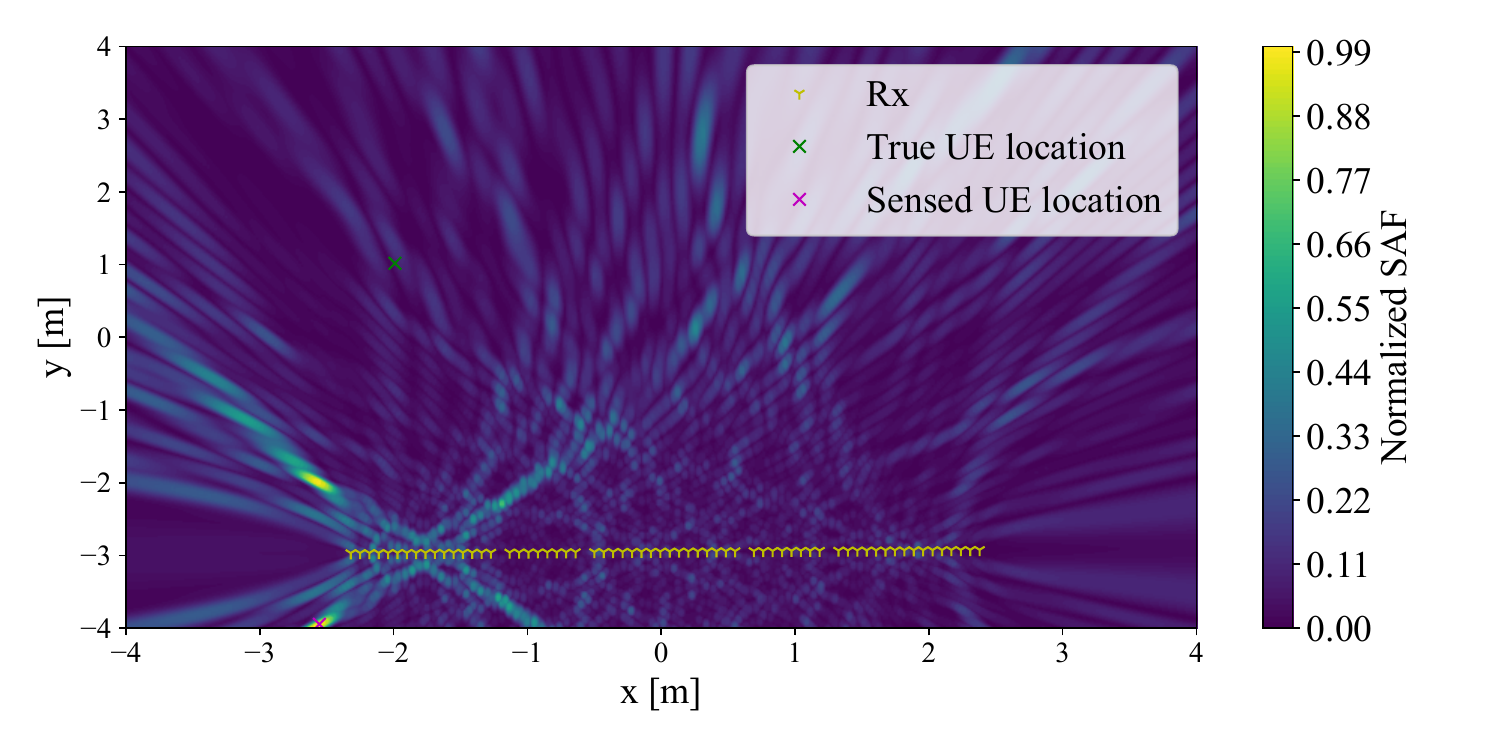} 
        \caption{}
    \end{subfigure}
    \begin{subfigure}[b]{0.32\linewidth}
        \centering
        \includegraphics[width=\linewidth]{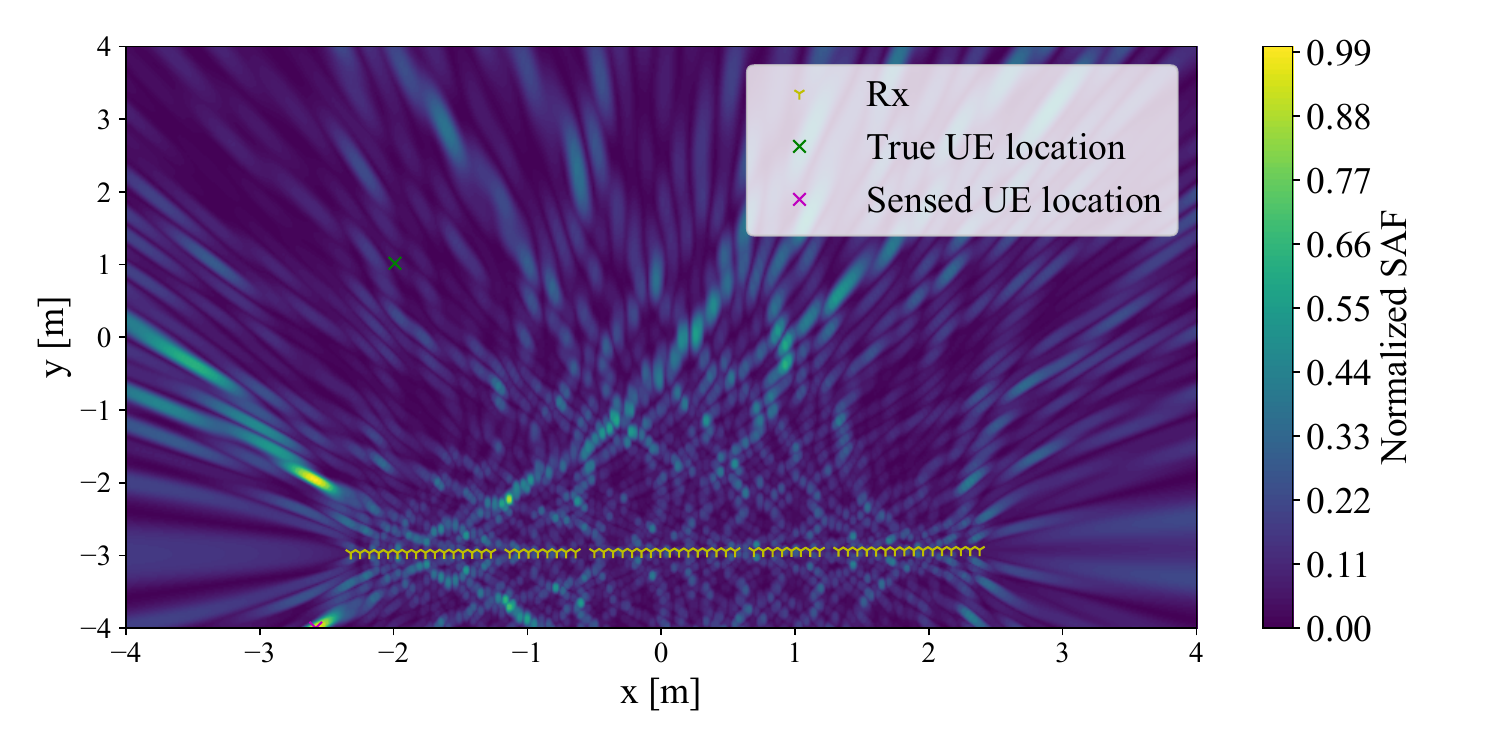}
        \caption{}
    \end{subfigure}
    \begin{subfigure}[b]{0.32\linewidth}
        \centering
        \includegraphics[width=\linewidth]{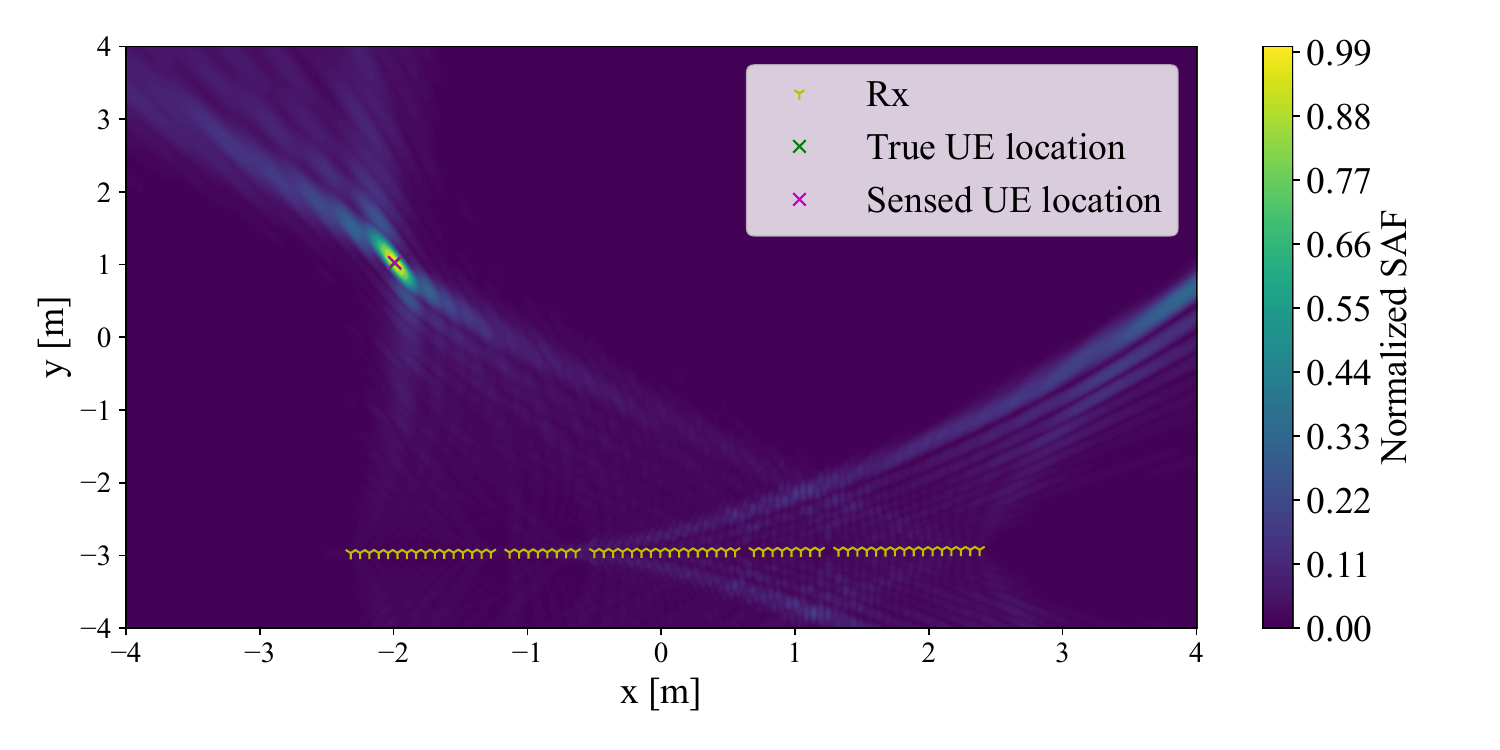}
        \caption{}
    \end{subfigure}

    \caption{\gls{ue} localization results: (a) Measured data. (b) After frequency phase offset compensation. (c) After both frequency and spatial offset compensations. \gls{pmsr} increases from \SI{14.01}{dB} to \SI{24.95}{dB}. Localization \gls{rmse} reduces from \SI{5}{m} to \SI{1.2}{cm}.}
    \label{fig: UE_localization}
\end{figure*}

\subsubsection{Localization Performance with Uncalibrated Data}

Fig.~\ref{fig: UE_localization}(a) shows the localization image obtained from uncalibrated \gls{csi} data. High sidelobe levels result in a low \gls{pmsr} of \SI{14.01}{dB}, making \gls{ue} localization difficult.



\subsubsection{Frequency Phase Offset Compensation Result}


Fig.~\ref{fig: UE_localization}(b) presents the \gls{ue} localization using the data after frequency phase offset compensation. Compared to Fig.~\ref{fig: UE_localization}(a), compensating frequency phase offset yields minimal improvement in localization performance, which aligns with the theoretical insights in Sections~\ref{subsec: crlb_result_frequency_offset} and~\ref{subsec: SAF_frequency_offset}.


\subsubsection{Spatial Phase Offset Compensation Result}

Fig.~\ref{fig: UE_localization}(c) shows the result after compensation for both frequency and spatial phase offsets. The sidelobes are substantially reduced, with \gls{pmsr} increasing to \SI{24.95}{dB}. The \gls{ue} can be accurately localized, indicating the effectiveness of spatial offset compensation, which also aligns with the simulation results in Sections~\ref{subsec: crlb_spatial_offset} and~\ref{subsec: SAF_spatial_offset}. Notably, the \gls{ue} localization \gls{rmse} reaches \SI{1.2}{cm} in the real-world scenario.


\section{Conclusion}
\label{sec: conclusion}
This paper presents a detailed study of the impact of phase incoherence on \gls{ofdm}-based massive \gls{mimo} systems for localization. We derived the \gls{crlb} and developed a unified model that uses the spatial ambiguity function to analyze the effects of frequency-dependent and (spatial) antenna-dependent phase offsets. Our simulation results demonstrated that, in the considered system configuration, frequency offsets have a limited effect on localization performance, whereas spatial offsets significantly degrade localization performance.

To mitigate these effects, we proposed a practical \gls{csi} calibration framework that targets both frequency and spatial phase distortions. The proposed method was experimentally validated on a real-world massive \gls{mimo} testbed, where we demonstrated its effectiveness in restoring phase coherence and improving localization. Specifically, the localization \gls{rmse} was reduced from \SI{5}{m} to \SI{1.2}{cm} after calibration, highlighting the critical role of spatial synchronization in localization.

Future work may extend the analytical simulations to wideband systems, where sensitivity to frequency offsets is higher. Additionally, the proposed calibration framework can be adapted for multi-\gls{ue} and passive target scenarios, providing robust phase synchronization in dynamic environments.

\bibliographystyle{IEEEtran}
\bibliography{ref}


\end{document}